\documentclass[thmsa,twocolumn,12pt,notitlepage]{article}
\usepackage{amssymb}

\begin{document}

\title{\textbf{The Quantum Universe}}
\author{\textbf{Jon Eakins and George Jaroszkiewicz} \\
School of Mathematical Sciences, University of Nottingham, UK}
\date{\textbf{Abstract: }\\
{\small \ On the basis that the universe is a closed quantum system with no
external observers, we propose a paradigm in which the universe jumps
through a series of stages. Each stage is defined by a quantum state, an
information content, and rules governing temporal evolution. Only some of
these rules are currently understood; we can calculate answers to quantum
questions, but we do not know why those questions have been asked in the
first place. In this paradigm, time is synonymous with the quantum process
of information extraction, rather than a label associated with a temporal
dimension. We discuss the implications for cosmology.}}
\maketitle

The exact nature of the universe is still uncertain at this time, with a
variety of incompatible paradigms, and models based on them, being assumed.
We believe that a unique paradigm for the universe can be found once certain
specific and reasonable answers to some standard questions are accepted.

The first important question is this: is the universe a complete quantum
system or not? By \emph{complete} we mean that there are no external
observers or classical agencies not subject to the laws of quantum mechanics.

The observation of Bell-type correlations and violations of Bell-type
inequalities together with a vast amount of empirical evidence for the
universal correctness of quantum principles strongly suggests that the
answer to the first question is \emph{yes.\ }This answer excludes from
further discussion all paradigms and models based on any sort of classical
hidden variables. We assume henceforth that the entire universe is a vast,
self-contained quantum automaton [1].

An immediate and unavoidable corollary following from this assumption is
that there are no semi-classical observers standing outside the universe or
any part of it. This seems at odds with the notion of observer in the
traditional formulation of quantum mechanics (also associated with the
so-called measurement problem), but should be reconciled with it in terms of
\emph{emergence}, i.e., the appearance of classicity from a fully quantum
universe.

From this corollary follows another unavoidable corollary, which is that the
universe must always be in a pure state, there being no global meaning to
the concept of a mixed state when discussing the entire universe. There can
be no intrinsic classical uncertainty as to which state the universe is in
at a given time. Mixed states can be discussed in our paradigm but only
under certain circumstances to do with emergence, when the notion of a
semi-classical observer can be invoked. This will be discussed elsewhere.

In a fully quantum universe, physicists who perform experiments and act as
observers with free will must themselves be quantum processes and part of
the very systems they appear to be monitoring. Accounting for them within a
pure state formulation is another aspect of emergence.\emph{\ }

Given that there are no external observers, the conclusion is that somehow
the universe organizes its own observation. This is not as strange as it may
seem at first sight. Sophisticated computers, for example, can run
diagnostic checks on themselves and repair problems in their software. In
the case of the universe, the question is: what does this observation mean?

The concepts of observation and measurement are meaningless without the
concept of \emph{time, }which is intrinsically associated with the process
of \emph{information extraction, }and it is only in these terms that we
understand\emph{\ }the dynamics of the universe.\emph{\ }In the standard
formulation of quantum mechanics, Schr\"{o}dinger evolution occurs precisely
in the absence of information extraction, and it is only when information
about a system is extracted that state reduction occurs. Therefore, we are
driven to the conclusion that time is no more nor less than a marker of a
constant process of state reduction in the universe which accompanies
information extraction. We differ from the multiverse paradigm on a number
of counts, the principal one being that we take state reduction as a
physically relevant process.

Given that time is a reflection of discontinuous quantum processes, then it
is at root a discrete phenomenon and not continuous. Therefore we may use
integers to represent time. According to our paradigm, at any given time $n$
the universe will be in a unique state $\Psi _n,$ the \emph{state of the
universe.} This state cannot be regarded as either a Schr\"{o}dinger picture
state or a Heisenberg picture state. Exactly what it is is bound up with the
concept of a \emph{stage,} discussed below.

The state $\Psi _n$ of the universe at a given time $n$ cannot by itself
represent the universe at that time in a complete way. In the conventional
formulation of quantum mechanics, for example, we would need to add to our
knowledge of the wave-function a statement of what the Hamiltonian was and
also, what sort of experiment or measurement we were going to perform. All
of this represents \emph{information }$I$ and \emph{rules }$\mathcal{R}$
(laws of physics), effectively telling us how the wave-function evolves.
This leads us to define the concept of a \emph{stage }as follows.\

To a given time $n$ we associate a unique stage $\Omega _n$ of the universe,
representing all possible attributes of the universe at that time. A stage
consists of three things: the current state of the universe $\Psi _n$, the
current information content $I_n$ of the universe, and the current rules $%
\mathcal{R}_n$ governing what happens next. We may write $\Omega _n=\Omega
\left( \Psi _n,I_n,\mathcal{R}_n\right) $ and assume that $\Psi _n$ is a
vector in some fundamental Hilbert space $\mathcal{H}$.

We refer now to Peres' discussion of the concept of quantum state
preparation, test and outcome [2] as the template for how the quantum
universe runs, meaning, how stages evolve.

First of all, contrary to Schr\"{o}dinger evolution, there can be no
deterministic sequence of stages. Given stage $\Omega _n$, we cannot in
general say with certainty what stage $\Omega _{n+1}$ will be. This is a
fundamental feature of our paradigm and marks the difference between the
notion of process time and block universe (or manifold) time. The principles
of quantum mechanics should apply at the level of the universe as well as
for subsystems. It is not that we cannot extract information about the
future. The future is not there until it happens and this is the correct way
to understand the uncertainty principle. The Kochen-Specker theorem [2]
supports Bohr's view that we are not interfering with a pre-existing
momentum when we measure particle position, and vice-versa. Neither
attribute of a particle system exists before we have specified the
experiment testing the state of the particle. This is perhaps the hardest
implication of quantum mechanics to accept.

The quantum dynamics of the universe runs as follows. At time $n$, the
current stage of the universe $\Omega _n$ contains enough information\ $I_n$
for the rules $\mathcal{R}_n$ to imply a generalized $test$ (in the sense
discussed by Peres [2]) $\Sigma _{n+1}$.\ This test is represented by some
self-adjoint operator $\hat{\Sigma}_{n+1}\;$in $\mathcal{H}$.\ It is not
clear at this time whether the rules should determine $\Sigma _{n+1}$
classically, that is, in an algorithmic, deterministic way, or whether $%
\Sigma _{n+1}$ is determined by (say) a form of quantum test itself. If $%
\Sigma _{n+1}$ is determined classically then it is some function $\Sigma $
of $\Omega _n$ and we may write $\Sigma _{n+1}=\Sigma \left( \Omega
_n\right) .$

The state of the universe $\Psi _{n+1}$ in the next stage is now postulated
to be one of the eigenstates of $\hat{\Sigma}_{n+1}$ with some eigenvalue $%
\lambda _{n+1}$ (or set of eigenvalues if $\hat{\Sigma}_{n+1}$ factorizes
into parts or if $\Sigma _{n+1}$ is represented by more than one operator).

The tests associated with stages will be extremely complex in principle,
because the Hilbert space $\mathcal{H}$ is expected to be vast. A simple
estimate of the minimum dimensionality of $\mathcal{H}$ may be based on the
idea of associating one qubit with each Planck volume in the visible
universe. This gives $\dim \mathcal{H}\gtrsim 2^{2\times 10^{180}}$for three
spatial dimensions and we expect it to be much bigger, if not infinite. In
principle the number of potential tests will be of the order ($\dim \mathcal{%
H)}^2$ [2]$,$ which is where the rules $\mathcal{R}_n$ come in. These will
effectively pick out of this vast number of potential tests just one, $%
\Sigma _{n+1},$ and then the universe will jump into one if its eigenstates.

The number of potential eigenstates may be vast, but one and only one of
these occurs when the Universe jumps from stage $\Omega _n$ to stage $\Omega
_{n+1}$, and this eigenstate is denoted by $\Psi _{n+1}$. The eigenvalues $%
\lambda _{n+1},$ test $\Sigma _{n+1},$ state $\Psi _{n+1}$and indeed $\Omega
_n$ are then convoluted with the old information content $I_n$ to form the
new information content $I_{n+1}$. Likewise, the outcome\ $\Psi _{n+1}$ and
new information content $I_{n+1}$ are convoluted with the old rules $%
\mathcal{R}_n$ to give a new set of rules $\mathcal{R}_{n+1}$, and then the
new stage $\Omega _{n+1}$ becomes the new reality. The old stage $\Omega _n$
now becomes unphysical and not even a memory. Any form of memory associated
with the past resides in the new information content $I_{n+1}$ alone. This
may include discrete topological information relating various factor states
in $\Psi _{n+1}$ and how they originated.

Which one of the possible eigenstates of $\hat{\Sigma}_{n+1}$ occurs cannot
be predicted, as this is the nature of the quantum process at work. However,
a probability estimate can be given, assuming that the computational
principles of quantum mechanics apply to the universe as well as to
subsystems. The conditional probability $P\left( \Psi _{n+1}=\Theta |\Psi
_n\right) $ that the state of the universe $\Psi _{n+1}$ at time $n+1$ is
eigenstate $\Theta $ of $\hat{\Sigma}_{n+1}$, given that the state at time $%
n $ is $\Psi _n,\;$is given by
\begin{equation}
P\left( \Psi _{n+1}=\Theta |\Psi _n\right) =|(\Theta ,\Psi _n)|^2,
\label{AAA}
\end{equation}
assuming that all potential outcome states are normalized to unity. This
probability estimate is valid only at time $n$, i.e. \emph{before }the jump,
and we may use it to discuss\ the potential future in terms of mixed states.
Once the universe has jumped from $\Omega _n$ to $\Omega _{n+1}$, however,
all probabilities alter and have to be revised.

The dynamics is such that total probability always sums to unity, but it is
not reversible. The apparent time reversal symmetry in $\left( \ref{AAA}%
\right) $ is more readily seen to be an artefact of the traditional Hilbert
space notation if instead we write $P\left( \Omega _{n+1}|\Omega _n\right)
\equiv |(\Psi _{n+1},\Psi _n)|^2.$ The direction of time is from $\Omega _n$
to $\Omega _{n+1}$ and not vice-versa. $\Omega _{n+1}$ is an outcome of test
$\Sigma _{n+1}$, which is determined by $\Omega _n$, whereas $\Omega _n$
cannot be an outcome of any test determined by $\Omega _{n+1}$. In fact, $%
\Omega _n$ was an outcome of some test $\Sigma _n$, which was determined by $%
\Omega _{n-1}$.

The direction of time is intrinsically that of information acquisition and
information loss. Stage $\Omega _{n+1}$ has information content $I_{n+1}$%
which need not contain $I_n$, and there is no algorithm which guarantees
that $\Omega _{n+1}$ could be used to deduce what $\Omega _n$ was, except in
the exceptional circumstance that the universe were reversible, which is
believed not to be the case on account of the second law of thermodynamics.

The stages paradigm permits further refinement relevant to cosmology. There
has been much speculation over the years concerning the discretization of
space and time on Planck scales. It has been pointed out by various authors
[3] that starting from discrete set models, the dimensions of emergent space
in a continuum limit may be dynamically determined by, for example, the
scale chosen. This raises the possibility that current speculations
concerning brane universe dynamics are consistent with the stages paradigm,
in that strings and branes, higher dimensional space and indeed general
relativity are but emergent approximations to a completely quantum universe.
The programme of quantizing gravity is from our perspective the wrong
direction to come, as has been suggested recently [4].

In the stages paradigm, the quantum process of test and outcome gives a
natural reason why time should be discrete. We might expect space also to be
discrete, but a straightforward discretization of emergent continuous space
would logically be the wrong thing to do without an analogous quantum
mechanism. In fact, a natural ``lattice'' structure does exist within the
stages paradigm. In any quantum theory, states in the Hilbert space may be
either fully entangled or products of factor states. In standard quantum
mechanics, factorization is a measure of \emph{classicity}.\ To describe a
state $\psi $ combining an identifiable apparatus and an identifiable system
under observation, we write\ $\psi =\Theta \otimes \varphi $, where $\Theta $
is the state of the apparatus and $\varphi $ is the state of the system.\
Without such a factorization, a classical distinction between apparatus and
system cannot be made. The origin of spatial discreteness may be indirectly
associated with the degree of factorization of the state $\Psi _n$ of the
universe at any given time. Information relating factor states in successive
stages gives a mechanism for the generation of ``family trees'', from which
causal set structures can arise, and ultimately, emergent space.

We propose the following schema for the cosmological history of the
universe. Assume that the Hilbert space $\mathcal{H}$ of the universe has
some enormous dimensionality, consisting of a direct product of a vast
number of fundamental qubits. Qubits are reasonable to assume because
fermions can be constructed from systems of qubits in the manner of Jordan
and Wigner [5] and fermions can be used to build up all types of particles.
In addition, qubits represent the most elementary yes/no kind of quantum
information, and information processing is at the heart of our paradigm.
Suppose further that the stages paradigm holds, and that the universe jumps
from stage to stage in the quantum way discussed above. Imagine now that the
state of the universe had once been completely entangled, i.e., consisted of
one factor, and had stayed like that over an enormous number of jumps.
During this epoch, which may be regarded as pre-big bang, there would have
been no possibility whatsoever of any classical structures such as space
emerging, and would truly have been a time of chaos in the original, ancient
world sense of the word.

It is not unreasonable to imagine that a non-zero probability existed that
the universe would sooner or later jump into a state which was the product
of two or more factors. Factorizable states belong to a subset of measure
zero in the set of all states, and therefore, a jump from complete
entanglement to a factorizable state would be highly improbable, except if
the rules $\mathcal{R}_n$ at a given stage made it more likely. The
empirical fact is that the current state of the universe does appear to be
factorizable, and so we may assume that this process occurred. Moreover, it
is possible to imagine that the rules and information content would then
change in such a way as to make the number of factors in succeeding stages
increase monotonically. This would signal the start of a quantum big bang,
and indirectly, the expansion of space.

In such a scenario, we may imagine that the onset of factorization was
followed by an epoch in which the number of factors in successive states $%
\Psi _n$ of the universe increased exponentially with jump time $n$, which
would be a period analogous to inflation. However, there need not be any
notion of Planck scale in this picture. The conventional Big Bang has to be
understood as something associated with emergence, which would occur only
long after the original quantum big bang. The individual factor states in
the factorizable state of the universe during inflation could themselves be
perhaps highly entangled states of vast numbers of qubits.

Further, entangled states can exhibit non-local correlations, the EPR
paradox being a good example. Such correlations within states in the early
universe might resolve the horizon problem in cosmology, accounting for the
observed isotropy and homogeneity of the current epoch.

A feature of this paradigm is that there is never any singularity associated
with the quantum big bang. The possibility arises of having many quantum big
bangs and quantum big crunches (a return to total entanglement), and also
many separate expanding universes, each regarded as an isolated set of
factors in the state of the universe. Each of these sub-universes would be
associated with its own emergent spacetime and there might be  no reason to
link them. If indeed the current information content $I_n$ contained no
information linking sub-universes at time $n$ then they would remain
separate ever after, there being no mechanism (by definition) for
re-entangling them at any future time. This is the quantum analogue of the
heat death of the universe.

The variable number of factors in the state $\Psi _n$ of the universe at
time $n$ gives a time dependent lattice\ $\Lambda _n$ called the \emph{%
factor lattice}. An element $\Psi _n^i$ of this lattice should not be
identified with any discrete point in physical space, as might be thought.
The relationship between the factor lattice and conventional spacetime is
more subtle on a number of counts, which we will discuss elsewhere. An
important feature here is that the factor lattice can easily encode the sort
of quantum non-locality observed in violations of Bell inequalities, and
this should manifest itself in the emergent limit in such a way that it did
not appear ubiquitous in classical spacetime, but only evident when looked
for in careful and sophisticated experiments.

Given that $\Psi _n=\Psi _n^1\otimes \Psi _n^2\otimes \ldots \otimes \Psi
_n^{N_n},$ where $N_n$ is the current number of factors, one way of defining
the \emph{current} \emph{classicity }$\kappa _n$ of the universe would be $%
\kappa _n\equiv \ln N_n/\ln N,$ where $N$ is the number
of qubits, assumed finite.
Then clearly\ $0\leqslant \kappa _n\leqslant 1.$ The lower bound zero
corresponds to a fully entangled, non-classical universe, whilst $\kappa
_n=1 $ corresponds to total classicity with no quantum entanglements
whatsoever. The current epoch of the universe appears to be one for which $%
\kappa _n$ is close to, but not yet, unity. The evidence for this is the
existence of a stable (and expanding) classical emergent space, but with a
residuum of quantum processes and correlations still active in the universe.

Given a high level of classicity, we may discuss quantum fields in terms
analogous to those used by Jordan and Wigner to construct fermions. Consider
some vast but finite subspace $\mathcal{H}^{\prime }\equiv \left\{ \mathcal{H%
}_1\otimes \mathcal{H}_2\otimes \ldots \otimes \mathcal{H}_{N_n}\right\} $
of factor qubit spaces in stage $\Omega _n$, which are not linked via the
information content $I_n$ to other factor qubit spaces in $\mathcal{H}$ at
time $n$. Then to all intents and purposes $\mathcal{H}^{\prime }$
represents some isolated sub-universe within the total universe associated
with $\mathcal{H}$.\ We may now construct fully anticommuting (fermionic)
quantum field operators acting on states in $\mathcal{H}^{\prime }$ in the
fashion of Jordan and Wigner [5], and thence construct relativistic quantum
field operators on the emergent energy-momentum space and emergent spacetime
associated with $\mathcal{H}^{\prime }$. The details will be given in
forthcoming papers.

Some final comments: the stages paradigm presupposes the existence of a
cosmic time in the universe, because quantum information acquisition is
inherently incompatible with closed timelike curves. However, this time is
``multi-fingered'', in that different factors in a given state $\Psi _n$ may
jump to new factors in $\Psi _{n+1}$ which are uncorrelated with each other,
and indeed, some factor states might not change from one stage to the next.
The overall picture is one of a quantum cellular automaton network structure
with ever changing topology predicated on quantum outcomes [1], leading to a
causal set structure [3]. The information content $I_n$ is a form of memory
which tracks over time the possibility of correlations between various
factor states, from which emergent structures such as continuous space with
a metric can be generated. The rules $\mathcal{R}_n$ are currently not
understood in any significant way. How they might change in time is related
to the question of how the laws of physics might change in time, and
understanding them is a challenge for the future, as is the difficult
problem of emergence.

Finally, the stages paradigm is based on the notion of process time.
Therefore, only one stage (known as the present) can be assumed certain in
any discussion. Relative to a given present, both the past and future do not
exist.

\

\textbf{Acknowledgement: }J.E. thanks the EPSRC (UK) for a research
studentship.


\begin{thebibliography}{9}
\bibitem{JAROSZKIEWICZ-01A}  G. Jaroszkiewicz, \emph{arXiv:quant-ph/0105013
v2}, pages 1--24 (2001).

\bibitem{PERES:93}  A. Peres, \emph{Quantum Theory: Concepts and Methods},
Kluwer Academic Publishers (1993).

\bibitem{SORKIN+al-87}  D. Meyer et al, \emph{Phys. Rev. Lett.},
59(5):521--524 (1987), G. Brightwell and R. Gregory, \emph{Phys. Rev. Lett.}%
, 66(3):260--263 (1991), M. Requardt, \emph{J. Phys. A: Math. Gen.},
31:7997--8021 (1998).

\bibitem{VISSER-01}  C. Barcelo and M.Visser, \emph{arXiv:gr-qc/0106002},
pages 1--5 (2001).

\bibitem{JORDAN+WIGNER-28}  P. Jordan and E. P. Wigner, \emph{Z. Physik},
47:631 (1928), J. D. Bjorken and S.D. Drell, \emph{Relativistic Quantum
Fields}, McGraw-Hill Inc (1965).
\end{thebibliography}
\end{document}